# Environmental qualification and performance of a CubeSat-scale dual optical frequency comb payload for space applications

**STEPHANIE D. LEIFER,[1,2,*] MATTHEW J. KELLEY,[1] NATALIE WALSH,[1] ZIBA SHAHRIARY,[1] PAUL STEINVURZEL,[1] CHRISSY FORTUNA,[1] GEOFFREY MAUL,[1] ADAM DARLEY,[1] IAN CODDINGTON,[3] LAURA C. SINCLAIR,[3] ARIEL BERMAN,[1] VICTOR CHIN,[1] HENRY TIMMERS,[4] KEVIN KNABE,[4] JASON PINON,[4] ANDREW ATTAR,[4] COLE SMITH,[4] CYRUS BRY,[1] DAVID S. KUN,[1] STAN STANEV,[1] NICOLE THAI,[1] BENNETT SODERGREN,[4] DANIEL DEVERAUX,[4] GRANT CAPAN,[4] WILL MCINTYRE ,[4] AND MANAS CHOURE[1]**

[1] *The Aerospace Corporation, 2310 E El Segundo Blvd, El Segundo, CA 90245, USA*
[2] *California Institute of Technology, 1200 E. California Blvd., Pasadena, CA 91125, USA*
[3] *National Institute of Standards and Technology, 325 Broadway, Boulder, Colorado 80305, USA*
[4] *Vescent Technologies, 14998 W. 6th Ave., Suite 700, Golden, Colorado 80401, USA*
**stephanie.leifer@aero.org*

**Abstract:** We report the environmental qualification and performance of CombSat, a dual optical frequency comb payload designed for deployment on a 12U CubeSat in low Earth orbit. The payload integrates two self-referenced, 200 MHz hybrid fiber-waveguide frequency combs, a narrow-linewidth optical reference, a heterodyne beat detection unit, and FPGA-based control electronics in a total mass of 8.3 kg and volume of 5.7 L, and nominal power consumption of 21W. We describe the mechanical, thermal, and radiation-tolerant design features of the system and present the results of random vibration, storage temperature cycling, and thermal-vacuum testing conducted in accordance with NASA's General Environmental Verification Standard (GEVS). We show that key comb performance metrics including oscillator and amplifier output power, $f_{rep}$, $f_{CEO}$ and $f_{opt}$ signal-to-noise ratios, and integrated phase noise, remain within specification after environmental exposure, and we discuss observed failure modes and design modifications that improved robustness. These results demonstrate that compact, low power, fully stabilized, autonomous dual-comb systems can be engineered and qualified for spaceflight on CubeSat platforms, providing a path forward for future low-cost comb-based space applications.

## 1. Introduction

Optical frequency combs are precise rulers for phase and frequency derived from frequency-stabilized mode-locked lasers. They provide a phase-coherent link between radio and microwave frequencies, underpinning a broad range of precision measurements, including absolute distance metrology [1-5], spectroscopy [6], quantum-limited optical time transfer [7-9], optical clock down-conversion for ultra-stable microwave generation [10,11], and optical atomic timekeeping [12]. Over the last two decades, frequency combs have been implemented in smaller packages with reduced power consumption [13,14], and have transitioned from laboratory settings to harsh terrestrial environments [15,16] and more recently, to space [17-22], with more flights anticipated within the next year [23-25]. However, to date, combs have not been deployed on CubeSat-scale spacecraft. Here, we describe the space environmental testing of CombSat, a CubeSat-scale dual frequency comb payload for deployment on a 12U CubeSat in low Earth orbit (LEO). Demonstration of such a low size, weight, and power (SWaP) payload will pave the way for regular space deployment in support of next-generation navigation and timing systems, high-precision tests of fundamental physics, and coherent optical links for deep-space communication. Optical time transfer at or below the $10^{-18}$ fractional level, combined with optical clocks on the ground or in orbit, would reset the standard for timing precision in space systems and could reduce the need for spaceborne advanced optical clocks.

CombSat integrates two self-referenced, 200 MHz hybrid fiber-waveguide frequency combs, a narrow-linewidth optical reference laser, a heterodyne beat detection unit (HBDU), and FPGA-based control electronics into a payload designed to emulate an autonomous spaceborne transceiver for sub-picosecond optical time and frequency transfer. However, the emphasis of this paper is on the system design choices and the environmental test campaign used to qualify the payload for a LEO mission. Details of payload performance for the optical time transfer application can be found in a companion paper [26].

Here we describe the mechanical, thermal, and radiation-tolerant design of the CombSat payload, including the use of some radiation-hardened components, a "sandwich" mechanical configuration compatible with a commercial 12U CubeSat bus, and a modular electronics architecture built around a Zynq-7030 System-on-Module. We then present the results of random vibration testing, storage temperature cycling, and thermal-vacuum (TVAC) testing conducted in accordance with NASA's General Environmental Verification Standard (GEVS) [27], as well as total ionizing dose (TID) testing of the waveguide module used for supercontinuum generation and *f-2f* self-referencing. Key comb performance metrics, including oscillator and amplifier output power, repetition rate ($f_{rep}$), carrier-envelope offset frequency ($f_{CEO}$) and optical beat ($f_{opt}$) signal-to-noise ratios (SNRs), and integrated phase noise, are compared before and after environmental exposure. We also document observed failure modes and the design modifications that mitigated them, which we believe may be of practical value to other groups preparing comb-based systems for space.

## 2. System Design

### 2.1 Payload architecture and SWaP

The CombSat payload is composed of four primary subsystems: (1) two complete, vacuum-compatible comb modules, each including pump diodes, a mode-locked fiber oscillator, optical amplifier, waveguide for supercontinuum generation and *f-2f* interferometry, and photodetectors for $f_{rep}$, $f_{CEO}$ and $f_{opt}$ generation, (2) a heterodyne beat detection unit that combines the amplified outputs of the two combs through a narrowband filter and balanced detector to generate dual-comb interferogram signals, (3) a narrow-linewidth (<1 kHz) optical reference laser that seeds both combs for optical frequency locking, an (4) the flight electronics for implementing comb stabilization, data acquisition, and payload control.

Figure 1 shows the comb modules and illustrates the payload assembly integrated within a Blue Canyon Technologies 12U CubeSat bus. The two comb packages are mounted face-to-face in a U-shaped bracket with the reference laser and HBDU optics located in the adjacent 2U volume next to the CubeSat's avionics module. The remaining ~4U of payload volume is reserved for potential co-payloads, such as an optical communications terminal [28] or a spectroscopy module for an optical atomic clock.

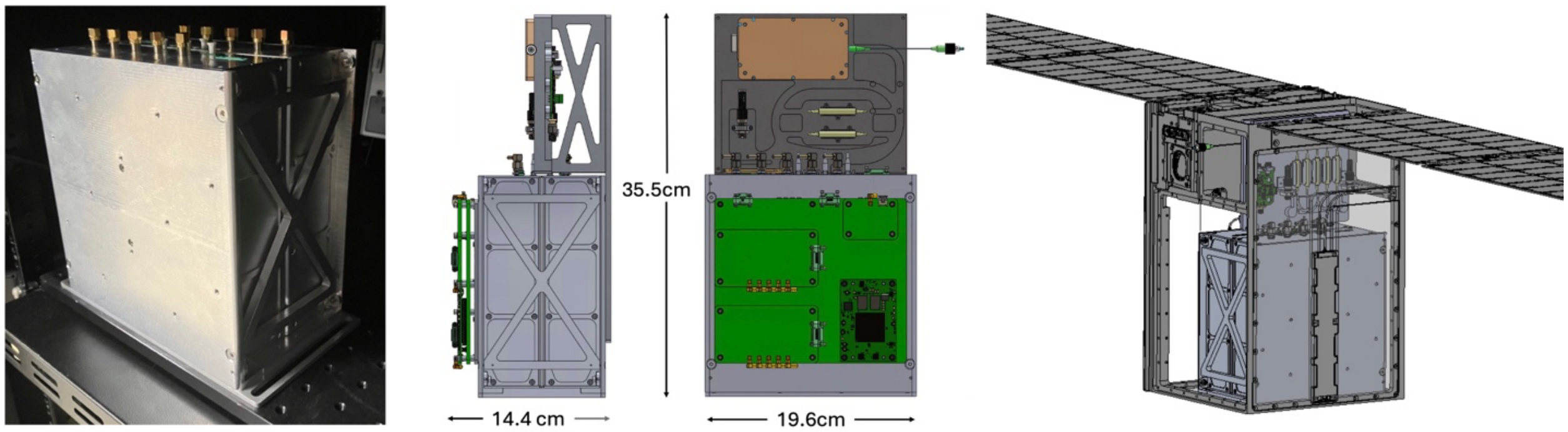

*Fig. 1. CombSat dual optical frequency comb payload. Left: Dual optical frequency comb flight development unit with comb packages mounted face-to-face in a U-shaped mounting bracket. Center: CAD model of the CombSat payload with the reference laser and HBDU optics located on the mounting bracket above the comb modules. The electronics boards are mounted in a stacked configuration to the inner-facing side of the U-shaped bracket housing the combs. Right: CAD model of the CombSat payload in a Blue Canyon Technologies 12U CubeSat bus. The 2U assembly in the upper left portion of the volume is the Blue Canyon Technologies XB1 avionics module. Image credit: Blue Canyon Technologies (bus), Vescent Technologies (comb modules), The Aerospace Corporation (all other payload components).*

Each comb module has a mass of 1.7 kg, occupies 1.93 L, and consumes 5 W under nominal conditions. The full CombSat payload, including both combs, optical reference, HBDU, and electronics, has a mass of 8.3 kg, occupies 5.7 L, and draws 21 W when the oscillator setpoint is near ambient temperature. At the temperature extremes of expected operation, additional thermo-electric cooler (TEC) power increases the total draw by up to 1 W per comb.

All fiber components in the comb modules are polarization-maintaining, and most optical components, including the oscillator fiber, are radiation hardened. There are no hermetically sealed components in the payload.

2.2 Optical frequency comb modules

Each comb module contains a self-referenced hybrid fiber-waveguide frequency comb operating at a nominal repetition rate of 200 MHz with spectrum centered near 1560 nm. The oscillators are passively mode-locked fiber lasers with cavity length controlled by a piezoelectric transducer (PZT) and TEC to tune $f_{rep}$. The output of the oscillator is amplified and injected into a tantala waveguide for supercontinuum generation and *f-2f* interferometry. The pump diodes (one each for the oscillator and amplifier) are integrated into the comb modules.

Each comb provides two fiber-coupled optical outputs: an oscillator tap with >10 μW of power and an amplifier tap with >10 mW integrated over 1555-1565 nm. Three RF outputs are generated: $f_{rep}$, the pulse repetition frequency (nominally 200 MHz), $f_{CEO}$, the carrier-envelope offset frequency derived from the *f-2f* beat, and $f_{opt}$, the heterodyne beat between the optical reference and the nearest comb tooth.

The oscillator cavity temperature can be set between 10 °C and 40 °C with a ±2.5 °C control range around the setpoint. The cavity PZT provides additional fine tuning of $f_{rep}$. $f_{CEO}$ is controlled via pump diode current and locked first; $f_{opt}$, which depends on both $f_{rep}$ and $f_{CEO}$, is locked second. The designed difference in repetition rates between the two combs is 2 kHz when both oscillators are at the same temperature.

2.3 Optical reference and heterodyne beat detection unit

A narrow-linewidth laser module at 192.115 THz serves as the optical reference for both combs. The module provides >10 mW output power with <1 kHz linewidth and includes an integrated thermo-electric cooler (TEC) controller and laser diode driver. Its output is attenuated and split via a 50:50 fiber coupler to deliver ~0.5 mW to each comb for generation of the $f_{opt}$ beat note. The optical reference establishes mutual coherence between the optical outputs of the two combs. Furthermore, the reference laser wavelength is close to the 1560 nm center of the output of the combs sent to the heterodyne beat detection unit (HBDU).

Thermal-vacuum testing of the reference laser module revealed an uncompensated wavelength shift of ~10 pm over a 10-50 °C temperature range, exceeding the ~0.5 pm tolerance required to keep the reference near a single comb tooth over the ± 100 Hz $f_{rep}$ tuning range. To mitigate this, we implemented a slow feedback loop that uses deviations of $f_{rep}$ from a target value referenced to a GPS-disciplined oscillator (GPSDO) to adjust the laser wavelength like Truong *et al.* reported [29]. With this loop engaged, the residual wavelength drift was reduced to 0.11 pm over the tested temperature range, corresponding to a residual $f_{rep}$ drift of ~±10 Hz, well within the combs' control capability [30].

The GPSDO therefore provides the absolute frequency reference for the combs. This choice minimizes system cost and complexity for the purposes of a space-based technology demonstration while maintaining mutual coherence between the combs due to the optical lock.

The HBDU combines the amplifier outputs of the two combs, filtered by a 13 nm band-pass filter centered at 1560 nm. Each comb output passes through a 0.03:99.97 fiber coupler. The 0.03% taps from the two combs are then combined in a 50:50 filter coupler and detected on a vacuum-tested 100 MHz balanced photodetector. The detector output is digitized at 200 MSPS by an ADC clocked from one comb's $f_{rep}$ signal.

2.4 Flight Electronics, Control, and Data

The CombSat flight control electronics depicted in Fig. 2 are comprised of a data processing unit (DPU), two analog boards, HBDU electronics, and the bus adapter. The system core is a Zynq-7030 based System-on-Module (SoM). The SoM is like those flown previously on International Space Station experiments [31,32]. A Zynq is also the core of the Red Pitaya-based comb control system widely used with laboratory optical frequency combs [33,34].

The DPU implements the command-and-control (C&C) interface to the combs through the analog boards, runs the phase lock control loops, and controls the main power conversion and distribution from the host bus. It also samples the output of the HBDU and controls the reference laser. The DPU implements 32 GB eMMC of nonvolatile memory for telemetry storage.

CombSat uses a modular design for the analog boards with one board per comb to reduce digital switching noise. Each analog board houses the data acquisition cards (DACs) to drive comb control signals, analog-to-digital converters (ADCs) to capture comb outputs, and a buffer for $f_{rep}$, which is used as the clock signal when locked. Each DAC and ADC have a scaling amplifier to optimally exercise the full range of the steering signals or capture the full range of the comb output signals.

The HBDU PCB integrates the balanced detector, a low-noise amplifier, and the 200 MSPS ADC. The ADC's sample clock is derived from one comb's $f_{\mathrm{rep}}$ signal and measures the beat note between the combs.

The CombSat payload was developed to be host-bus agnostic but requires an adapter board to convert the power and C&C interfaces from any number of spacecraft platforms to be compatible. The main high-speed C&C interface to the bus is 1Gb Ethernet via a serial gigabit media independent interface (SGMII). There is also an auxiliary low-speed 3.3V universal asynchronous receiver/transmitter (UART) interface. CombSat is designed for 12V, 5V, and 3.3V power inputs.

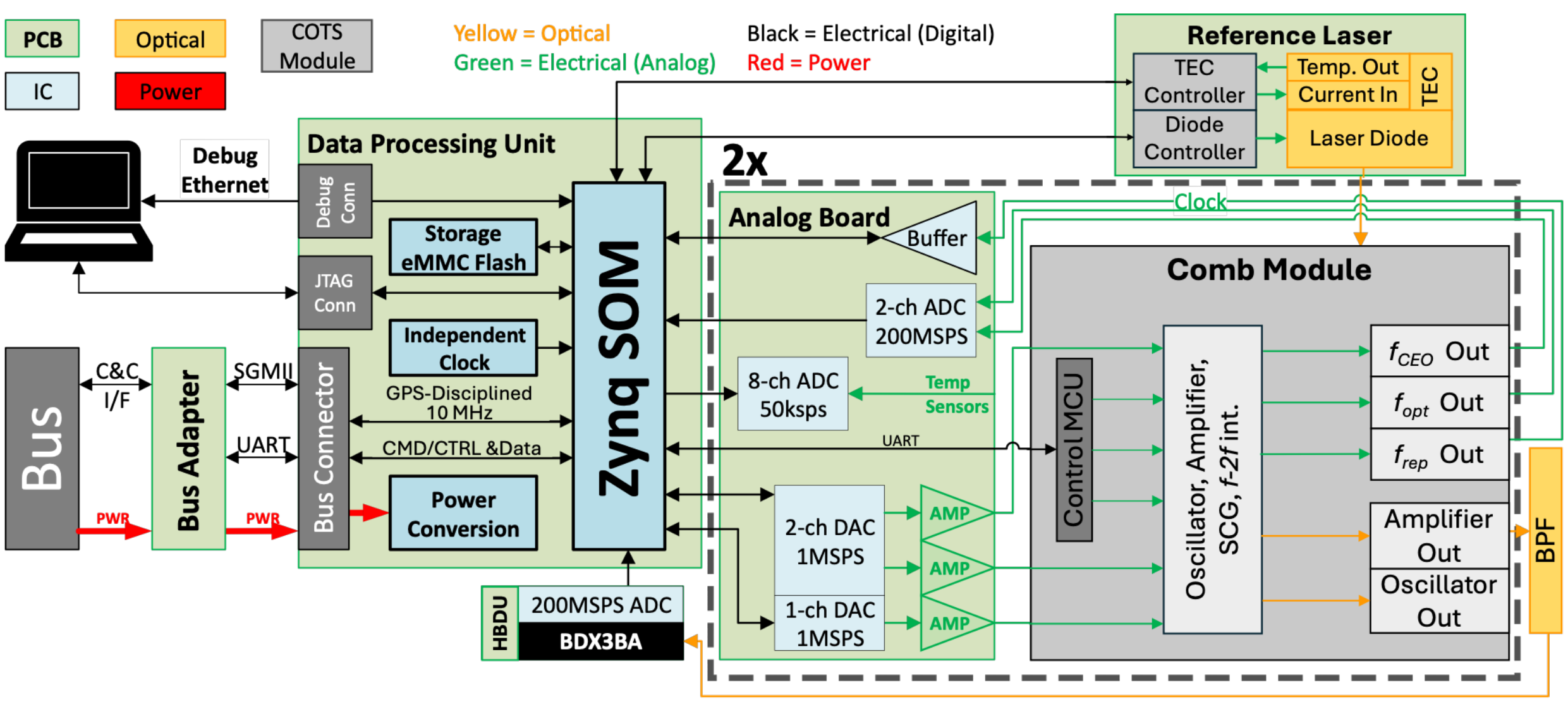

*Fig. 2. The CombSat flight electronics architecture. The design is comprised of three PCBs: a Data Processing Unit (DPU) built around a Zynq SoM, two Analog boards to interface with the combs, and the heterodyne beat detection unit (HBDU) card. BPF=Band Pass Filter, MCU= Microcontroller unit, SCG= supercontinuum generator, TEC=thermoelectric cooler.*

The electronics were first demonstrated in a large form-factor breadboard version composed of three rack-mounted chassis. This breadboard enabled component-level evaluation and test, firmware and software integration, and end-to-end testing with the combs before miniaturization into a flight-like payload. CombSat's system software and firmware architecture and anticipated data volume description can be found in the Supplementary Material.

2.5 Mechanical configuration

The CombSat payload is designed to be compatible with a Blue Canyon Technologies (BCT) 12U CubeSat bus. The two comb modules are mounted face-to-face in a U-bracket ("sandwich" configuration) as shown in Fig. 1. This configuration minimizes thermal gradients between the oscillators and allows both combs to be maintained at similar temperatures with minimal TEC power.

The bracket includes helicoil mounting holes on the bottom and sides for attachment to the bus structure, and additional fasteners are inserted internally on the top of each comb chassis to suppress relative motion under vibration. Small pilot through-holes are machined into all mounting holes on the bottom of the bracket to provide vent paths and avoid trapped volumes during ascent and in vacuum.

The 2U volume in the upper left-hand corner of the image represents BCT's XB1 avionics module. The 2U volume adjacent to the XB1 contains the reference laser and HBDU optical components. The mounting plate for these components and the fiber routing were engineered to fit within this volume to preserve more space (nearly 4U) for an additional payload.

## 3. Environmental Considerations – Modeling and Test

Environmental testing of the dual comb package was conducted in accordance with the guidelines in the NASA General Environmental Verification Standard (GEVS) [27]. The test sequence was vibration, followed by storage temperature (thermal soak) testing, and completed with thermal vacuum. Table 1 summarizes the results of these tests.

Each parameter value measured prior to environmental testing is shown for both combs along with the required value range. The post-vibration-testing parameter value change (Δ), and additional post-thermal soak change are noted in the adjacent columns, followed by the values measured during thermal vacuum exposure at the nominal base plate temperature of 10 °C, and the extreme values of -5 °C and +25 °C. In the following sections, we detail each test sequence and related modeling.

**Table 1. Comb performance pre- and post-vibration and storage temperature testing, and during thermal vacuum environmental testing.**

| Parameter | Units | Spec. | Value | | | | | | | | Notes |
|---|---|---|---|---|---|---|---|---|---|---|---|
| | | | Comb 1 | | | | Comb 2 | | | | |
| | | | Pre Test Value | Post vibe Δ | Post Soak Δ | TVAC** @-5°C/10°C /25°C | Pre Test Value | Post vibe Δ | Post Soak Δ | TVAC** @-5°C/10°C /25°C | |
| $f_{rep}$,Pulse Repetition Rate | MHz | 200 | | | | 213.8/194/192.7 | | | | 214.4/195/193.6 | ± 1kHz, 20 °C Oscillator setpoint, Combs not locked |
| Oscillator Output Power | μW | >10 | 21.3 | 0 | 1 (4.7%) | 21.3/20.9/20.0 | 17.7 | -0.1 (0.5%) | 1.9 | 17.7/17.3/16.5 | |
| Amplifier Output Power | mW | >5 | 30.1 | 0 | 2.6 (8.6%) | 28.7/30.1/30.1 | 20.5* | 8.2 (40%) | 4.1 (14.3%) | 28.7/27.1/28.8 | |
| $f_{CEO}$ RF Signal SNR | dB | >40 | 41 | 0 | 1 (12.2%) | 42.5/42.2/42.0 | 48.1 | -2.1 (-21.5%) | -3 (-29.2%) | 42.6/40.0/43.2 | @100kHz RBW, 10kHz VBW |
| $f_{CEO}$ RF Signal Integrated Phase Noise | mrad | <6000 | 610 | 0 | 50 (8.2%) | 710/690/710 | 660 | 0 | -70 (-10.6%) | 630/600/620 | In-Loop, 10 Hz – 1 MHz |
| $f_{opt}$ RF Signal SNR | dB | >40 | 55 | 0 | -2 (-20.6%) | 53.0/50.7/50.5 | 59 | -12 (-74.9%) | 8 (151.2%) | 52.9/54.6/54.5 | @100kHz RBW, 10kHz VBW |
| $f_{opt}$ RF Signal Integrated Phase Noise*** | mrad | | 190 | 0 | 30 (15.8%) | | 180 | 0 | 70 (38.8%) | | 10 Hz – 1 MHz |
| $f_{rep}$ SNR | dB | >50 | 67 | 0 | 1 (12.2%) | 73/75/75.5 | 69 | -1 (-10.9%) | 4 (48.5%) | 72.8/76.0/74.0 | @100kHz RBW, 10kHz VBW |
| Maximum Power for 2 Combs | W | <20 | | | | | 12/11.6 | | | | Baseplate: -5°C/25°C Oscillator: 10°C /20°C |

**The Comb 2 amplifier diode current before vibration testing was 460mA, and after the test was 310mA. This was changed because it was determined that prior to the test, the amplifier was operating in a regime that was not optimized for $f_{CEO}$ generation, leading to significant sensitivity to changes in pump current. To avoid conflating the change in $f_{CEO}$ due to vibration testing with a change due to nonoptimal operating currents, the pump current was adjusted.*
***In TVAC, the oscillator TEC was changed to 10°C for the -5°C baseplate case because the TVAC baseplate temperature could not be maintained at -5 °C when the oscillator was set to 20 °C.*
**** Not specified*

## 3.1 Thermal

### 3.1.1 Thermal analyses

Thermal analyses of the CombSat payload were performed using the Thermal Desktop software. The engineering laboratory combs were modeled, then actual thermal data obtained in the laboratory for validation. The validated model was then imported into a bus model. Next, analyses were executed for beta angles (the angle between the spacecraft's orbital plane and the Sun vector) from 0° through 75° in a 470 km circular orbit for both hot and cold extreme cases (set by solar irradiance, Earth infrared irradiance, and albedo) to determine the payload temperature variations throughout an orbit. These analyses drove the power budget estimate by determining the current needed for each comb's thermo-electric cooler (TEC) and advised the target operating temperature setpoint to minimize total power draw. Figure 3a shows the Thermal Desktop-modeled assembly. Figure 3b shows the locations on the mounting brackets where the temperatures reported in Figs. 3c and 3d were determined. Figures 3c and 3d show the predicted Comb 1 oscillator temperature variation throughout an orbit and the TEC current required to hold the oscillator temperature to 20ºC, respectively. Figures 3e and 3f show the same for Comb 2. The 20ºC oscillator setpoint temperature was chosen because it corresponds to a mounting plate temperature of 10ºC which was found to be the most likely value (mode) for beta angles between 0º and 45º. The cases shown are for both the hot and cold extremes for beta angles of 0°, 45°, and 75°. In all cases, all system components can be kept within their design temperature operating range.

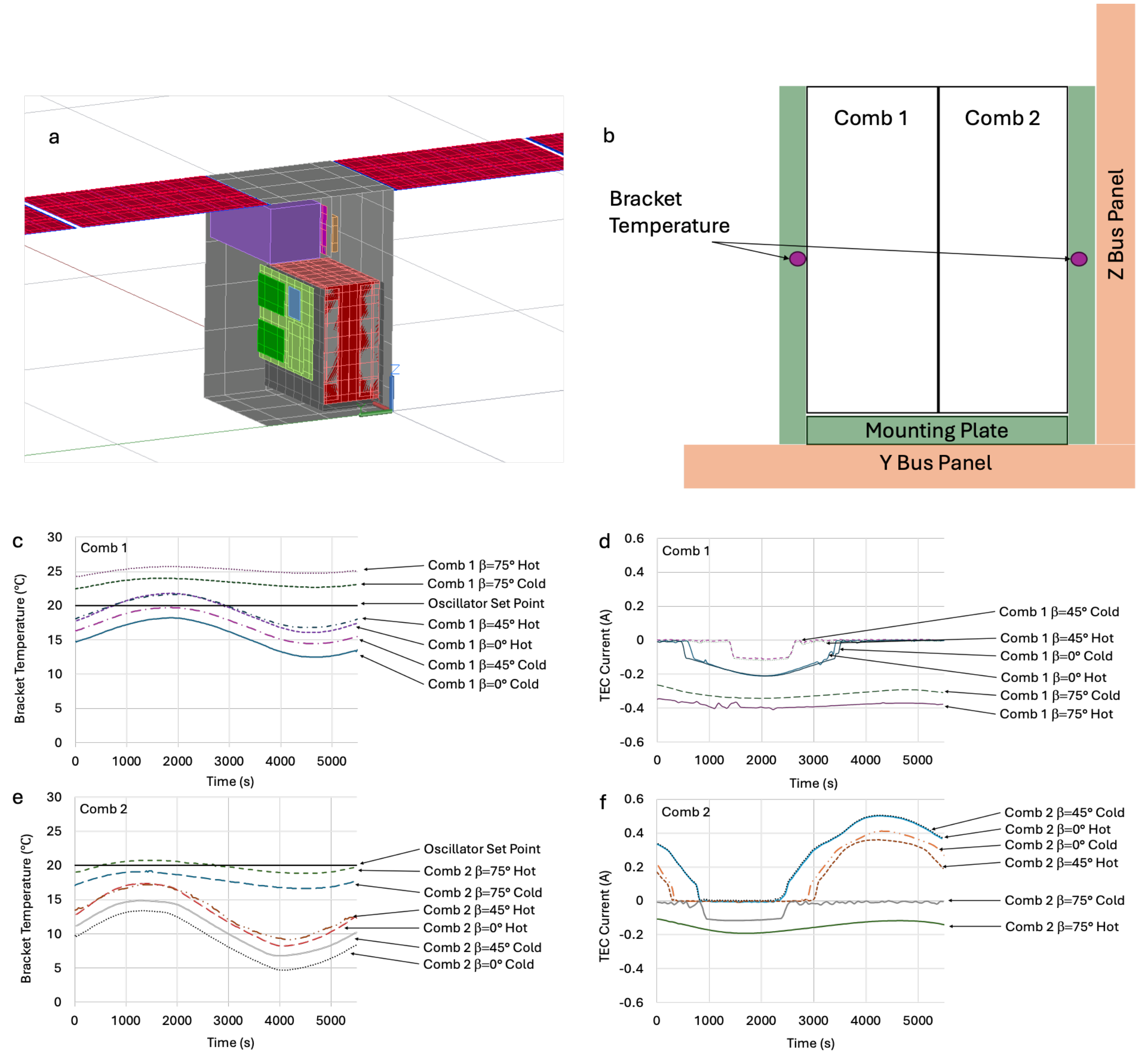

*Fig. 3. CombSat thermal analyses. (a) Thermal Desktop model of the CombSat payload in a Blue Canyon Technologies 12U CubeSat bus in a 470 km Earth orbit. (b) Bracket locations where the temperature determined in the model is reported. (c) Bracket temperatures for Comb1 over the course of an orbit when the oscillator TECs are held to 20℃. Results for beta angles of 0°, 45° and 75° for both hot and cold cases are shown. (d) TEC current over an orbit required to maintain the 20℃ temperature for the Comb 1 oscillator. (e) and (f): Same as (c) and (d), but for Comb 2.*

### 3.1.2 Storage Temperature Testing

Storage temperature testing was conducted by cycling the non-operating dual comb package from -5°C to 40°C for a soak time of 12 hours at each temperature with a ramp rate between temperatures of 1.5 hours. The combs were tested before and after thermal soak to ensure no degradation of critical operating parameters. Performance parameters are shown in Table 1. We note that storage temperature testing had the most pronounced effect on the difference between pre- and post-test parameters of all the tests performed. This is attributed to two effects. First, thermal expansion and contraction likely caused slight changes in alignment and mechanical stress. This behavior is observed in other combs where, after the initial storage temperature cycle, subsequent temperature cycles show reduced changes in performance metrics. Second, we have observed that humidity affects fibers with acrylate coatings and presumably induces strain. Thus, at the high temperature setpoint, the buffers may have been drying out. Regardless, all parameters remained within the required performance range.

### 3.1.3 Thermal Vacuum Testing

Thermal vacuum (TVAC) testing of the dual comb assembly was performed in a chamber at a pressure <$7\times10^{-3}$ Pa. Figure 4a shows the dual combs in the vacuum chamber. The optics modules were mounted on a thermally controlled baseplate that was variable between -5ºC and 25ºC such that the dual comb assembly could achieve the temperature range expected on orbit per the thermal model. Both the optical reference laser and a FPGA controller for comb locking were used on the exterior of the vacuum chamber.

The CombSat mechanical assemblies are designed to survive the effects of rapid depressurization during ascent at levels below 0.65 psi/sec. Venting adequacy was established by satisfying the empirical rule V/A < 2000 inches where V = the total internal "void" volume of the assembly in cubic inches and A is the total area of the vent paths in square inches. These guidelines are in accordance with accepted standards [35].

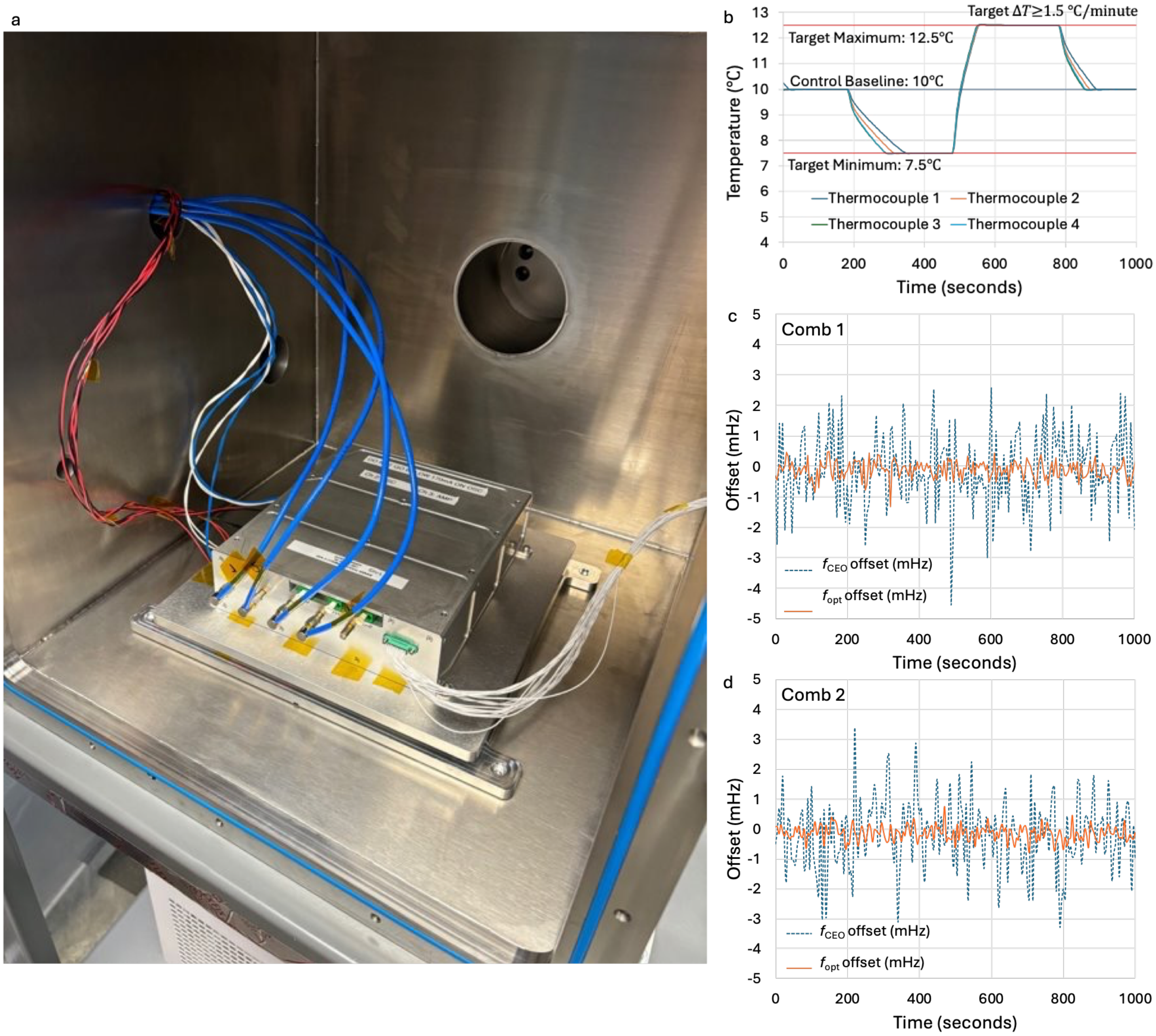

*Fig. 4. CombSat thermal vacuum chamber testing. (a) Comb modules in thermal vacuum test chamber. The units demonstrated successful operation, generating the required $f_{rep}$, $f_{CEO}$ and $f_{opt}$ control signals while under vacuum over a baseplate temperature range of -5℃ to 25℃. (b) Profile of the temperature ramp applied to the baseplate while the combs were locked to the reference laser at an oscillator temperature setpoint of 20℃. The baseplate temperature was measured at four different locations in separate quadrants to ensure an even temperature distribution. (c) and (d) $f_{opt}$ and $f_{CEO}$ signal offsets from 21.4 MHz setpoints of both locked combs during the ramp.*

In the first set of tests, the oscillator temperature was set to 20ºC on each free-running comb and performance was evaluated over a baseplate temperature range from 7.5ºC to 12.5ºC with a ramp rate of 1.5 C/min to simulate orbital temperature changes. The temperature profile is shown in Fig. 4b and reflects the temperatures reported at four separate thermocouple locations on the baseplate. Phase noise of $f_{\mathrm{CEO}}$, SNR of $f_{\mathrm{CEO}}$ and $f_{\mathrm{opt}}$, SNR and frequency of $f_{rep}$, optical output power of the oscillator and amplifier taps, and total power consumption were all monitored and are reported in Table 1. During the temperature ramp, the free-running drift of $f_{rep}$ was measured to be much less than 1 kHz offset from nominal for both combs (0.034 kHz for Comb 1 and 0.023 kHz for Comb 2) over the full temperature range.

To evaluate performance at temperature extremes beyond the expected worst-case, the baseplate temperate was varied between -5ºC and 25ºC. When the baseplate temperature was 25ºC, the oscillator temperature of each comb remained at the 20ºC temperature setpoint, and all the aforementioned parameters except $f_{rep}$ were monitored for 30 minutes. The oscillator temperature was set to 10ºC when the baseplate was set to -5ºC. This was due to the limitation of the TVAC chamber to remove heat from the baseplate, and not from a limitation of the comb temperature control itself. The soak duration at the temperature extrema prior to testing was 1 hour each.

Once the combs were under vacuum, it was found that a -5 kHz to -6 kHz shift from the target $f_{rep}$ value of 200 MHz had been induced, even under constant temperature. While a reduction in the mean refractive index of the optical cavities would be expected to increase the repetition rate, other factors, including changes in mechanical stress and humidity due to outgassing, had a larger impact. The offset can be accounted for in future builds. However, there was an additional ~1 kHz $f_{rep}$ shift resulting from a baseplate temperature change from 10°C to 25°C, even when the oscillator temperature setpoint was held fixed at 20°C. We believe the additional shift is due to changes in strain of the optical fiber given the thermal gradients over the full oscillator package; the 20°C TEC setpoint was measured at only one location. Nevertheless, there was ample tuning range for $f_{rep}$ in both combs to compensate for this effect -- 30 kHz when changing the oscillator temperature from 20°C to 40°C at a fixed baseplate temperature of 10°C.

At the -5°C baseplate setpoint, the $f_{rep}$ offset from 200 MHz was +14 kHz due to the oscillator setpoint temperature being changed to 10°C. At this low temperature setpoint, the oscillator temperature was ramped between 0°C and 15°C, corresponding to a 25 kHz tuning range.

All other measured parameters met the specifications shown in Table 1. The maximum combined comb power consumption over the baseplate temperature ramp between 7.5°C and 12.5°C was 10.75W. At -5°C (and a 10°C oscillator setpoint) it was 12W and at 25°C (20°C oscillator setpoint) it was 11.9W.

The combs were independently locked while under vacuum with the external laser serving as the frequency reference. The frequency stability of $f_{\mathrm{opt}}$ and $f_{\mathrm{CEO}}$ were measured during the temperature ramp profile shown in Fig. 4b. In-loop $f_{\mathrm{CEO}}$ and $f_{\mathrm{opt}}$ were shown to be stable throughout the ramp for each comb (Figs. 4c and 4d).

## 3.2 Launch loads

### 3.2.1 Finite Element Analysis

Finite Element Analysis (FEA) was performed on the dual comb assembly using the Ansys Mechanical software package where the Finite Element Model (FEM) was created from solid models of the combs and detailed material tables. Each comb module FEM was built up from five distinct sub-groups of components, then a top-level U-bracket FEM was set up with the two comb modules mounted inside, representing the complete assembly. Twenty-two constraint points were set using the mounting hole locations on the side and bottom of the bracket. Lateral acceleration was applied to both (x, z) axes, and all 3 accelerations were run simultaneously. Figure 5 shows the FEM assembly von-Mises stress and deflection results, which indicated positive margins on the dual comb assembly for the 10 G axial and 17 G lateral loads expected during launch. The minimum margin of safety value found was 1.3 relative to yield on a commercial, off-the-shelf standoff on the comb assembly housing. This exceeds the required factor of safety specified in NASA-STD-5001B [36], indicating that the structure will not yield under the design load conditions. The margin implies that in a future development, reduced mass is possible.

### 3.2.2 Random Vibration Testing

Random vibration testing was performed on the unpowered combs to assess their survivability using the protoflight test vibration profile defined by NASA GEVS corresponding to 14.1 $G_{\mathrm{rms}}$ over the frequency range of 20 Hz to 2 kHz. The test was conducted on the x-axis configuration shown in Fig. 6a for 60 seconds using a Vibration Research VR4600 electrodynamic shaker and controller and VibrationVIEW software.

Sine sweep testing was conducted at a sweep rate of 4 octave/min from 3-2000Hz and an amplitude of 0.5 $G_{rms}$. The sweep was performed both before and after random vibration testing to ensure the combs have no natural frequencies lower than 100 Hz and therefore are less susceptible to resonances. Comb performance was also measured both before and after vibration testing.

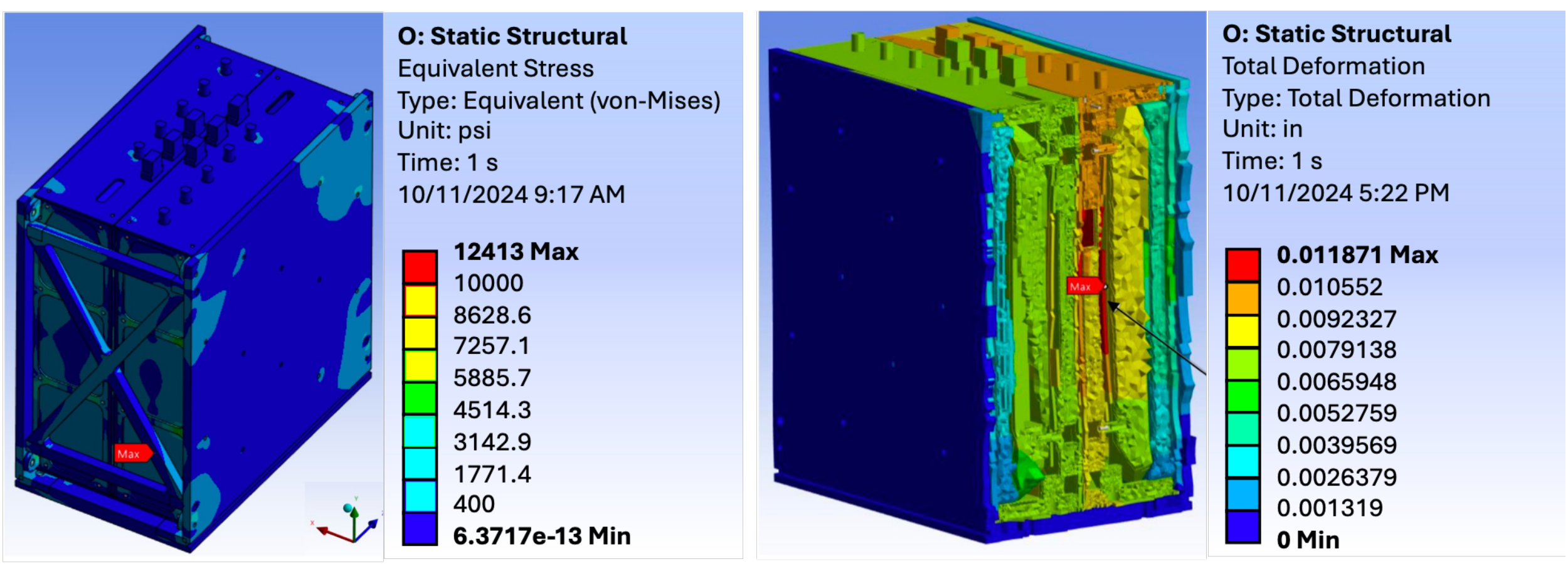

*Fig. 5. Finite element model of the CombSat assembly showing von-Mises stress (left) and deformation results (right). The results show positive margins on the dual comb assembly for 10 G axial and 17 G lateral loads expected during launch. The minimum Margin of Safety (MoS) value found was 1.3 on a commercial, off-the-shelf standoff on the comb assembly housing.*

Figure 6b shows the low-level sine-sweep acceleration profiles for all three axes of the comb package. No sub-100 Hz resonances were observed. However, multiple design issues were discovered after an initial round of random vibration testing (see applied random vibration power spectral density in Fig. 6c) and required system modifications to prevent future issues.

Both modules were damaged; there were broken fibers on the routing plate between the tantala waveguide module and the comb module lid. This was subsequently redesigned for better fiber retention and avoidance of pinch points. The other failure mechanism was a shift in the optimum amplifier current of the $f_{\mathrm{CEO}}$ beat note caused by a sensitive fiber splice that shifted the gain fiber, causing poor coupling of the common port of the amplifier wavelength division multiplexer (WDM). A switch was made to a different WDM, and this was shown to be less sensitive. We note that many WDM modules contain free-space optics, and this is an important consideration for flight.

After repairs were made and the units retested, Comb 1 behaved normally with no failures, but Comb 2 had a faulty communication issue due to a flat flex cable connection between the oscillator and the comb PCB that became unseated during vibration. In response, the flat flex connector with replaced with a Harwin gecko connector for a more secure connection.

Also, the input collimator on the tantala waveguide module in Comb 2 broke. The top screws that typically secure the lid on the module were removed for vacuum compatibility, leaving only the screws on side of the module to affix the lid. This created a lever arm that vibrated during testing, resulting in epoxy inside the module severing the input collimator and resulting in no visible $f_{\mathrm{CEO}}$ signal. The waveguide module was replaced and a new design developed.

## 3.3 Radiation

### 3.3.1 Electronics

Because CombSat is designed to fly in Low Earth Orbit (LEO), non-radiation-hardened electronic components were selected to minimize cost. However, for passive components, only those that met Automotive Electronics Council AEC-Q200 global stress test qualification standards were chosen. Specifically, we use only surge current tested, non-electrolytic capacitors with high dielectric ratings (100V, 50V) and tantalum or tantalum polymer capacitors where higher capacitance than available in ceramics is needed. For transistors, diodes and MOSFETs, we use Joint Army-Navy Test Extended (JANTX) components for flight. AEC-Q100 ICs and AEC-Q101 discrete semiconductors are preferred for the engineering development model for reduced cost. To improve overall radiation tolerance, we partition the power architecture to isolate failures and prevent cascade. We also monitor currents and voltages to quickly initiate a shutdown in case of irregularities. For processor and memory components we utilize Error Correction Code (ECC) memory and configuration memory scrubbing for increased resilience to single event functional interrupts and rely on heartbeat messages, echo commands, and packet counting techniques to indicate system health and allow reset of hung-up systems.

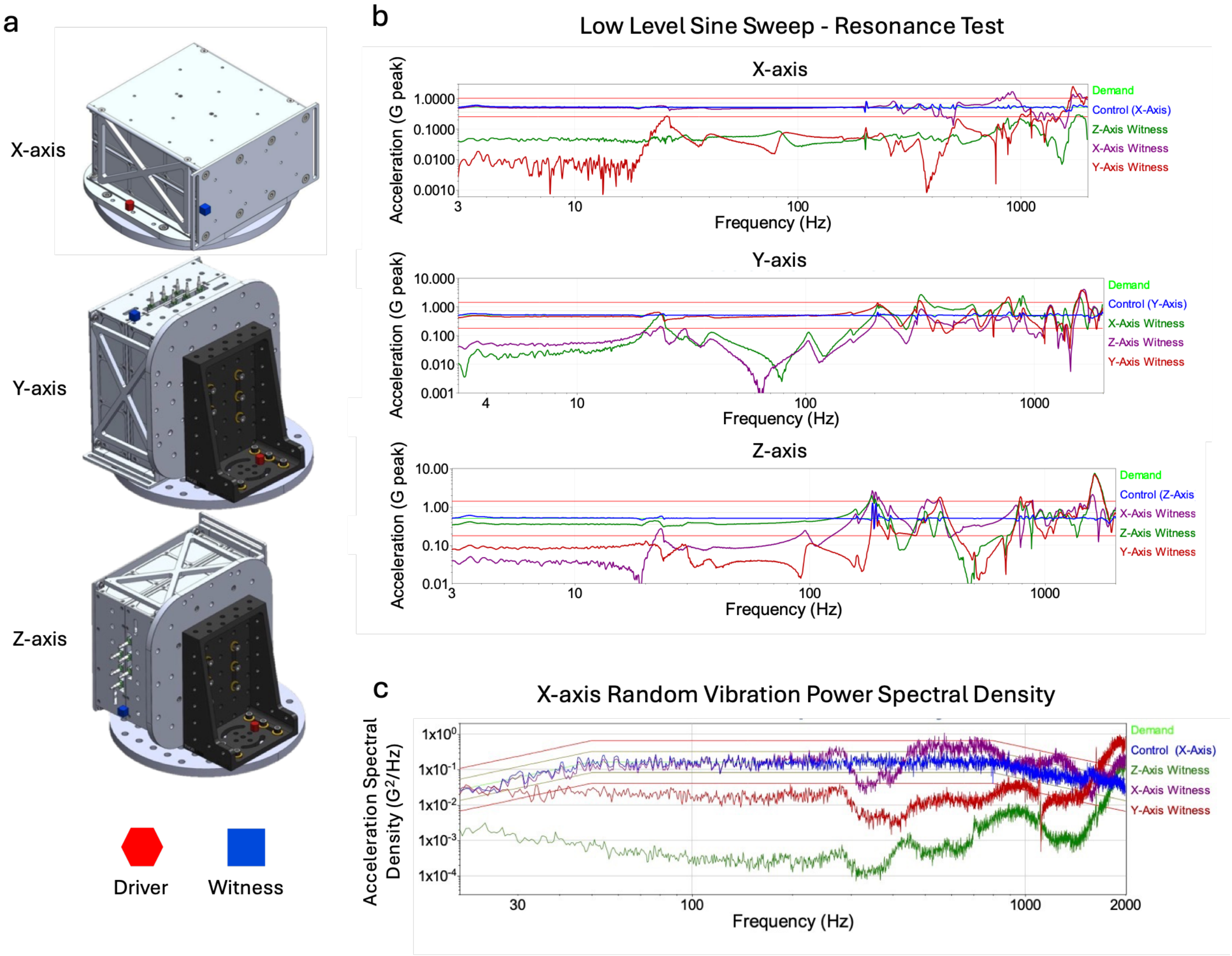

*Fig. 6. Acceleration spectral density from low-level sine sweeps on each axis of the dual comb assembly. In each plot, the control signal for the measured axis is shown in dark blue. The measured witness acceleration profiles for each of the other axes are plotted and labeled according to the legend on the right side of each plot. Note that there were no observed resonances below 100 Hz. There was a small change in shape to the resonance observed at ~900 Hz, but this magnitude of change is within the measurement-to-measurement variation. (a) mounting configurations and locations of the driver and witness, (b) Acceleration profile for x-, y-, and z axis mounted comb modules for low-level sine-sweep, (c) random vibration profile for the x-mounted configuration.*

### 3.1.2 Optical components

As mentioned previously, the combs are built with radiation hardened optical fiber, including the oscillator cavity. As a precautionary measure to prevent loss of modelocking, this results in a slightly narrower tuning range than is found for a similar comb built with standard optical fiber.

The supercontinuum-generating tantala waveguide (COSMO module by Octave Photonics) was radiation tested in a J. L. Shepherd Model 484 tunnel irradiator using γ-rays emitted by $^{60}$Co. The dose rate was measured to be 39 rad(Si)/s before irradiation. Input optical power was applied throughout exposure to 3 Mrad(Si) and the unit was biased to operate normally. The RF spectrum at the output was continuously monitored *in-situ*. The COSMO module demonstrated no statistically significant effects due to total ionizing dose (TID), maintaining the same SNR of the $f_{CEO}$ signal throughout exposure [37].

While the LEO orbit intended for CombSat does not present a challenging radiation environment, these results illustrate CombSat's optical system design extensibility to higher orbits.

## 4. Conclusions

We have developed CombSat - a space-capable, self-referenced, low size, weight, and power dual optical frequency comb payload appropriate for future space-based applications on a 12U CubeSat-scale platform in low Earth orbit. Vibrational and thermal environmental testing demonstrated that the dual comb system successfully met all environmental requirements after critical design modifications. Specifically, we demonstrated that the combs can survive launch loads and maintain lock during the full range of thermal variations expected on orbit while meeting all performance requirements. Furthermore, the CombSat architecture is extensible to other frequency comb designs and to higher orbits. An orbital flight demonstration of CombSat is the logical next step in proliferation of low-SWaP space-based frequency comb applications.

### *Supplementary Material*

The CombSat high-level software architecture, algorithm for on-orbit operation, and anticipated data volume is provided in the supplementary material.

### *Funding*

The CombSat project was funded by The Aerospace Corporation's Innovation Laboratory and The Defense Systems Group, Communications and Navigation Capabilities Division; Position, Navigation, and Timing Futures Space Acquisitions Office, and by the United State Space Force Space Systems Command, Office of the Chief Technologist.

### *Acknowledgment*

The CombSat team thanks Mr. Charles Player of the Aerospace Corporation for supporting the CombSat program, Jamil Abo-Shaeer and Arman Cingoz of Vector Atomic (now part of IonQ, Incorporated) for their laboratory combs and technical support, and the Position, Navigation, and Timing Group of the Air Force Research Laboratories at Kirtland Air Force Base, and Benjamin Stuhl of the Space Dynamics Laboratory for useful discussions. We are grateful to Blue Canyon Technologies for drawings and parameters of their 12U CubeSat. Thanks also to Mahmood Bagheri of NASA's Jet Propulsion Laboratory for equipment loans early in the program development.

### *Disclosures*

The authors declare no conflicts of interest. The use of tradenames in this manuscript is necessary to specify experimental results and does not imply endorsement by the National Institute of Standards and Technology nor the Aerospace Corporation. During manuscript preparation, the authors used Astro, an internal AI assistant provided by the Aerospace Corporation, to suggest wording changes and improve clarity in the text. No AI tools were used to generate scientific content, data, or analysis. All AI-assisted suggestions were reviewed, edited, and approved by the authors, who take full responsibility for the manuscript.

### *Author Contributions*

SDL conceived of and led the project and prepared the manuscript. MJK, IC, AB, and DSK contributed to the manuscript. MJK led the laboratory testing, and CB, and MC contributed to laboratory testing. IC and LCS consulted on frequency comb requirements specifications, PS wrote the flight comb requirements specification, CF and GM did the mechanical design, SDL, MJK, CB, MC, and LS developed the algorithms for the software team, DSK, NT, and SS developed the flight software and firmware, AB and VC designed and the flight electronics PBC boards and AB, VC, and SS assembled flight emulator electronics. NW performed the flight system thermal analysis, AD was the flight systems engineer, and ZS managed project budget and schedule, and advised on shock and vibration qualification testing. HT developed the rad-hard oscillator design and down-stream fiber-optic design, KK adapted a previous comb design into the LEO-qualification CombSat design, JP did the mechanical design of the combs and performed thermal simulations, AA was the dual comb systems engineering and project manager, BS contributed to the optomechanical and fiber-optic design and process engineering, and CF and BS selected epoxies, DD designed and commissioned the TVAC experiments, GC developed the comb firmware and software ICD, WM did the comb PCB electronics design and layout, and CS built the dual-comb system and performed conformance testing.

### *Data Availability*

The data that support the findings of this study are available within the article.